\documentclass[aps,prb,reprint,amsmath,amssymb,floatfix,hidelinks]{revtex4-1}
\usepackage{graphicx}
\usepackage{dcolumn}
\usepackage{bm}
\usepackage{hyperref}
\usepackage[T1]{fontenc}
\usepackage[latin9]{inputenc}
\usepackage{caption}
\usepackage{amsmath}
\usepackage{amssymb}
\usepackage{epstopdf}
\usepackage[caption=false]{subfig}
\usepackage{color}
\usepackage{natbib}

\makeatletter

\@ifundefined{textcolor}{}
{%
 \definecolor{BLACK}{gray}{0}
 \definecolor{WHITE}{gray}{1}
 \definecolor{RED}{rgb}{1,0,0}
 \definecolor{GREEN}{rgb}{0,1,0}
 \definecolor{BLUE}{rgb}{0,0,1}
 \definecolor{CYAN}{cmyk}{1,0,0,0}
 \definecolor{MAGENTA}{cmyk}{0,1,0,0}
 \definecolor{YELLOW}{cmyk}{0,0,1,0}
}

\usepackage[english]{babel}

\begin{document}

\title{Emergence of the stripe-domain phase in patterned Permalloy films}

\author{S. Voltan$^1$, C. Cirillo$^2$, H. J. Snijders$^1$, K. Lahabi$^1$, A. Garc\'{i}a-Santiago$^{3,4}$, J. M. Hern\'{a}ndez$^{3,4}$, C. Attanasio$^2$, J. Aarts$^1$}

\affiliation{$^{1}$Kamerlingh Onnes-Huygens Laboratory, Leiden University, P.O.
Box 9504, 2300 RA Leiden, The Netherlands.\\
$^{2}$CNR-SPIN Salerno and Dipartimento di Fisica "E.R. Caianiello", Universit\`{a} degli Studi di Salerno, via Giovanni Paolo II 132, I-84084 Fisciano (Sa), Italy.\\
$^{3}$Grup de Magnetisme, Departament de F\'{i}sica Fondamental, Facultat de F\'{i}sica, Universitat de Barcelona, c. Mart\'{i} i Franqu\`{e}s 1, planta 4, edifici nou, 08028 Barcelona, Spain.\\
$^{4}$Institut de Nanoci\`{e}ncia i Nanotecnologia IN2UB, Universitat de Barcelona, c. Mart\'{i} i Franqu\`{e}s 1, planta 3, edifici nou, 08028 Barcelona, Spain.}

\begin{abstract}
The occurrence of stripe domains in ferromagnetic Permalloy (Py=Fe$_{20}$Ni$_{80}$) is a well known phenomenon which
has been extensively observed and characterized. This peculiar magnetic configuration appears only in films with a
thickness above a critical value ($d_{cr}$), which is strongly determined by the sputtering conditions (i.e. deposition
rate, temperature, magnetic field). So far, $d_{cr}$ has usually been presented as the boundary between the homogeneous
(H) and stripe-domains (SD) regime, respectively below and above $d_{cr}$. In this work we study the transition from
the H to the SD regime in thin films and microstructured bridges of Py with different thicknesses. We find there is an
intermediate regime, over a quite significant thickness range below d$_{cr}$, which is signaled in confined structures
by a quickly changing domain-wall configuration and by a broadening of the magnetoresistance dip at the coercive field.
We call this the emerging stripe-domains (ESD) regime. The transition from the ESD to the SD regime is accompanied by a
sharp increase of the magnetoresistance ratio at the thickness where stripes appear in MFM.
\end{abstract}
\date{\today}

\maketitle
\selectlanguage{english}%

\section{Introduction}\label{secin}
Alloys of iron and nickel, known as permalloys, are much exploited in applications because of their particular magnetic
properties. In particular, Permalloy with approximately 20\% Fe and 80\% Ni (Py=Fe$_{20}$Ni$_{80}$) is widely used in
magnetoelectronic devices such as, for example, magnetic recording media, magnetic transducers, MRAM and magnetic cores
of inductors~\cite{amos,khizroev,dastagir,zhao}. At this specific composition the values of magnetostriction and
magnetocrystalline anisotropy are nearly zero. As a result, Py is characterized by a very high permeability
($\mu_r\simeq 8000$) and low coercive field (below 1 mT), which makes it a ``soft'' ferromagnet.

\begin{figure}[ht]
\begin{center}
\includegraphics[width=0.9\linewidth]{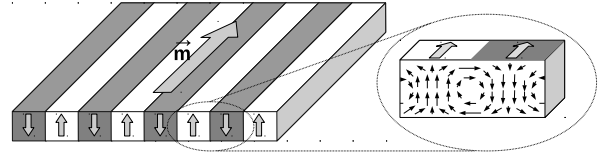}
\caption{Sketch of the magnetization directions in the weak stripe domain phase of a film of Permalloy. The main magnetization direction
is given by $\vec{m}$. An alternating perpendicular is shown as gray and white areas. The zoom highlights the direction of the out-of-plane component and
the variation of this component in a domain wall.}\label{sketch}
\end{center}
\end{figure}

In Py thin films, because of the demagnetizing field, the magnetization normally lies in-plane. However, if grown under
particular conditions, Py films can have a certain amount of Perpendicular Magnetic Anisotropy (PMA). This leads to the
occurrence of magnetic stripe domains (SDs) \cite{saito,magn_domains}. If the PMA is small, as in the case of Py, the
domain state is called ``weak stripes'': the main magnetization component is still in the film plane but it is tilted
alternatively upwards and downwards by a small out-of-plane component\cite{magn_domains}, as sketched in
Fig.~\ref{sketch}.
SDs appear only above a certain value of the film thickness, given by $d_{cr} = 2 \pi \sqrt{A / K_{\bot}}$, where $A$
is the exchange stiffness constant and $K_{\bot}$ is the perpendicular anisotropy constant. SDs in Py have been
experimentally well characterized and their peculiar properties have been exploited in magnetic devices for several
purposes\cite{dastagir, belkin, vlasko}. However, to our knowledge, little work has been done to describe the
transition regime below $d_{cr}$. Micromagnetic simulations were performed to investigate the type of domain walls
occurring in narrow strips~(Ref.~\cite{magn_domains}, Ch.3.6) as function of thickness and perpendicular anisotropy,
but those results do not signal the changes we observe with increasing thickness.

As a matter of fact, values for $d_{cr}$ are hardly ever determined, nor quantitatively compared to values extracted
for $K_{\bot}$ from e.g. magnetization measurements. Instead, studies on stripe domains are simply performed on films
with thicknesses well above an inferred critical thickness. Our interest in the magnetic structure of Py films derives
from studies of superconducting Py/Nb multilayers, in which unusual behavior was found of the superconducting critical
fields. For relatively large thicknesses (of the order of 200 nm) but {\it below} the onset of stripe domains we find
indications for a long range proximity effect and the occurrence of odd-frequency triplet Cooper pairs, which appears
to be the consequence of an inhomogeneous magnetic state in the ferromagnetic layer\cite{berg_2001, esch_phystod}. A
discussion of the proximity effects will be given elsewhere, but it also led us to a systematic study of the magnetic
behavior of our Py films as function of thickness, using Magnetic Force Microscopy (MFM), ferromagnetic resonance
(FMR), SQUID magnetometry and magnetoresistance measurements (MR). We study the behavior of full films as well as of
confined structures, such as bars and squares, and we also use micromagnetic simulations to compare with experimental
results. For confined geometries, the results show that the influence of the perpendicular anisotropy can be found well
below the stripe domain (SD) regime, in particular in the structure of domains and domain walls, and the behavior of
the MR. This leads us to identify two different regimes below the SD regime: a fully homogeneous (H) regime for thin
samples, and a regime which we call emerging stripe-domains (ESD), for intermediate thicknesses. In the ESD regime the
perpendicular anisotropy clearly influences the magnetic configuration even if without forming full stripes. In the
description we use an operational definition of $d_{cr}$ as the thickness where stripes appear in MFM measurements,
which in our case is around 300 nm. We show that this coincides with a strong increase in the magnitude of the
magnetoresistance dip around the coercive field. On the other hand, the MR data display a decided broadening of the dip
in the regime between 150~nm and 300~nm (ESD). Thus, we argue that the homogeneous magnetic state already disappears at
less than 0.5 $d_{cr}$. We also discuss how this picture is to be reconciled with out in-plane and out-of-plane
magnetization data.

The paper is organized as follows: in Sec.\ref{secexp}, we describe sample preparation details, measurement and
simulation methods; in Sec.\ref{secmfm} we describe MFM images of Py films and structures in the three different
thickness regimes, in Sec.\ref{secmag}, we present FMR and magnetization vs field measurements of Py films, used to
determine the critical thickness, while in Sec.\ref{secmr} we show and discuss the magnetoresistance measurements;
micromagnetic simulations of confined structures are presented in Sec.\ref{secsim} and, to conclude, Sec.\ref{seccon}
highlights the main results of our study.

\section{Experimental details and methods}\label{secexp}
Py films were deposited on Si(100) substrates in a ultrahigh vacuum DC diode magnetron sputtering system, at room
temperature. The base pressure reached was approximately 2.7$\times10^{-8}$ mbar, while the deposition was done in an
Ar pressure of 2.7$\times10^{-3}$ mbar. The typical deposition rate, measured by a calibrated crystal monitor, was
0.30~nm/s. Several series of Py films with different thickness were prepared, called S1 (50, 200 and 350~nm), S1b
(290~nm), S2 (50, 100, 150, 200, 250, and 360~nm), S3 (100, 125, 150, 175, 200, 225, 250, 275, 300, 325 and 350~nm) and
S3b (380~nm). The growth conditions were nominally the same for all samples, but they were grown at different times.
The samples of the same series were grown in succession, within one or two days. Magnetic imaging was both performed on
as-grown films and on films patterned into small structures via e-beam lithography followed by Ar-ion etching. The
structures were small squares, as well as bridges with contacts in standard 4-probe geometry (current contacts outside,
voltage contacts inside) for the transport measurements. For all devices on which transport measurements were made, the
width of the bridge was 10~$\mu$m and the distance between the voltage contacts 100~$\mu$m.

Magnetic imaging was performed on both unpatterned (S1 series) and patterned samples (125~nm and 225 from the S3
series, 380~nm from the S3b series) with standard Magnetic Force Microscopy (MFM), in lift height mode. Magnetic
hysteresis loops of unpatterned samples from the S2 series were taken with a commercial (Quantum Design) SQUID
magnetometer, while the broadband microstrip FMR\cite{denysenkov} was performed on the unpatterned samples of the S2
series. An Agilent E8361A PNA Millimeter Wave Vector Network Analyzer (10MHz-67GHz) was used to apply a microwave
signal to the samples and to measure the magnetic absorption. The signal is injected into a microstrip line on top of
which the sample is located. We register the complex microwave scattering parameter S$_{21}$ as a measure of the
microwave magnetic absorption. The FMR responses for all samples were measured at room temperature by sweeping the
frequency for fixed external applied field in the 0.25-15 GHz range. This process was repeated for several applied
field values ranging form -50~mT to 50~mT.

The electrical measurements were done with an automated measurement platform (PPMS), with the magnetic field applied
in-plane and along the current direction, on the samples of the S3 series (except for the 225~nm thickness). We need to
point out that the magnetoresistance curves presented (Fig.\ref{mr1}) are affected by a systematic offset along the
$x$-axis (up to 20~mT), which is positive for backward sweeps and negative for forward sweeps, and dependent on the
starting field value. Because of that, the dip in the MR curve occurs {\it before} the field reaches zero value. This
error, introduced by the magnet remanence (of PPMS) at low fields, is more extensively discussed in the Supplementary
Information. The offset becomes a problem when determining the exact coercive field; however it does not influence the
discussion below, for which only the MR ratio and the dip width are relevant.

Micromagnetic simulations were performed with the software package \textsc{oommf}\cite{oommf} (object oriented micromagnetic
framework) for square structures $4 \times 4~\mu$m$^2$ and thickness in the different regimes (100, 225, 285 and
345~nm). The cell size used for the calculations is $8 \times 8 \times 15~$nm$^3$ or smaller and the damping
coefficient is 0.5. The details for the magnetic parameters used are presented in Sec.\ref{secsim}.

\section{Experimental results and Discussion}

\subsection{Magnetic Force Microscopy}\label{secmfm}
\begin{figure}[ht]
  \centering
  \includegraphics[width=0.75\linewidth]{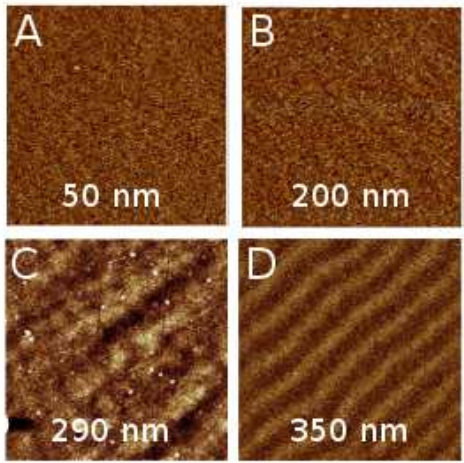}
   \caption{\label{mfm1} (Color online) Phase contrast images from Magnetic Force Microscopy for full films in the three different thickness
   regimes: A: 50~nm, B: 200~nm, C: 290~nm and D: 350~nm. A,B and D are from the S1 series, C from S1b. Scan areas are $5\times5\ \mu$m$^2$.}
\end{figure}
MFM images of unpatterned Py films with thicknesses in the three different regimes from the S1 series (50~nm, 200~nm
and 350~nm) and S1b (290~nm) are shown in Fig.\ref{mfm1}. For the thickest sample, clear stripe domains are observed
(Fig.\ref{mfm1}-D). Darker and brighter regions (domains) represent areas where an out-of-plane component of the
magnetization is detected and points upwards or downwards, respectively. The domain width is approximately 330~nm,
which is of the same order of the thickness of the sample, as predicted for weak stripe domains\cite{magn_domains}. No
contrast is observed for the samples 50~nm (Fig.\ref{mfm1}-A) and 200~nm thick (Fig.\ref{mfm1}-B), which suggests that
either the magnetization is fully in-plane or the out-of-plane component is below the sensitivity of our MFM detection.
For the 290~nm thick sample (Fig.\ref{mfm1}-C) we can observe non-homogeneous magnetic areas, even if they are not
fully developed in stripes yet. Given these observations, the critical thickness $d_{cr}$ for our samples can be
defined to be slightly above 300~nm.

\begin{figure}[ht]
  \centering
  \includegraphics[width=8cm]{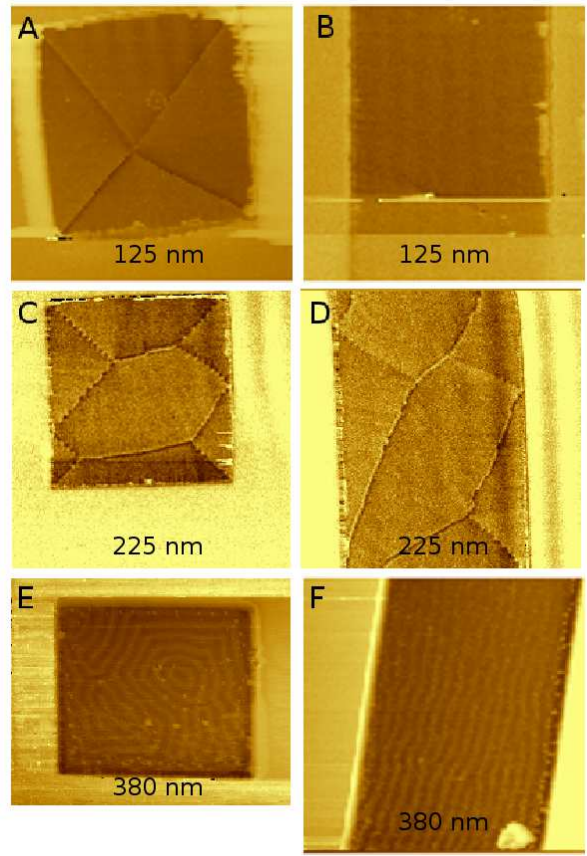}
   \caption{\label{mfm2} (Color online) Phase contrast images from Magnetic Force Microscopy for patterned Py in the three different thickness regimes.
   A,B: 125~nm; C,D: 225~nm from the S3 series; and D,E: 380~nm from the S3b series. Structures in A and C are squares $10\times10\
   \mu$m$^2$, E is $10\times9\ \mu$m$^2$. In B,D and F a portion of a 10 $\mu$m wide bar is shown; scan ranges are $15\times15\
   \mu$m$^2$ for A,C,D and F, $14\times14\ \mu$m$^2$ for (B) and $15\times12\ \mu$m$^2$ for E.}
\end{figure}
To further investigate the magnetic configuration at different thicknesses, MFM measurements were performed also on
patterned films of the S3 and S3b series, in particular on squares of approximately $10\times10\ \mu$m$^2$ (see
Fig.\ref{mfm2}-A,C,E) and long bars of 10~$\mu$m wide (see Fig.\ref{mfm2}-B,D,F), in the three different regimes.
Fig.\ref{mfm2}-E and -F show that for samples 380~nm thick (S3b series), well above $d_{cr}$, the confinement does not
hinder the presence of stripe domains. In Fig.\ref{mfm2}-E the effects of the demagnetizing fields lead to rotations of
the stripe directions, producing maze-like domain configurations. Also in Fig.\ref{mfm2}-F the stripes are clearly
visible and they are aligned along the bar, parallel to the magnetic field previously applied to magnetize the virgin
sample. In this case the stripes turn out to be stronger in the proximity of the extremity of the bar (the edge is just
outside the scan range) and they become weaker while moving far away from it. The reason is that at the center of the
bar the shape anisotropy forces the magnetization to be more in-plane, weakening the out-of-plane component. At the
extremity, instead, the influence of the shape anisotropy is weaker and the stripes are less affected. Fig.\ref{mfm2}-A
shows a structure of 125~nm thickness. Here, as we expect, the magnetization is fully in-plane, so the magnetic
configuration is mainly determined by the demagnetizing energy, which results in four triangular closure domains, with
Bloch domain walls. For the 225~nm thick sample shown in Fig.\ref{mfm2}-C we observe a magnetic configuration which is
in between the other two regimes: there are triangular closure domains and a large center domain where the
magnetization is fully in-plane and no stripes are visible; most of the domain walls now seem to be ``broken'', with
alternating up-down components, indicating that the out-of-plane anisotropy is playing a role, even if it is not strong
enough to generate stripes. Such a difference is clearly visible also for the bars, as can be observed by comparing
Fig.\ref{mfm2}-B (125~nm thick) and Fig.\ref{mfm2}-D (225~nm). In the first case triangular domains, similar to the
ones observed in Fig.\ref{mfm2}-A, are confined to the extremities (not shown in the image), but the magnetization is
homogeneously in-plane in the rest of the structure. In the second case (Fig.\ref{mfm2}-D), the domains are present in
the whole bar with the characteristic configuration observed also in Fig.\ref{mfm2}-C.

\subsection{FMR and Magnetometry}\label{secmag}
As discussed in the Introduction, the critical thickness $d_{cr}$ can be, in principle, determined by estimating the
uniaxial (weak) out-of-plane anisotropy $K_{\bot}$ and the exchange constant $A$.

\begin{figure}[ht]
    \centering
    \includegraphics[width=1.0\linewidth]{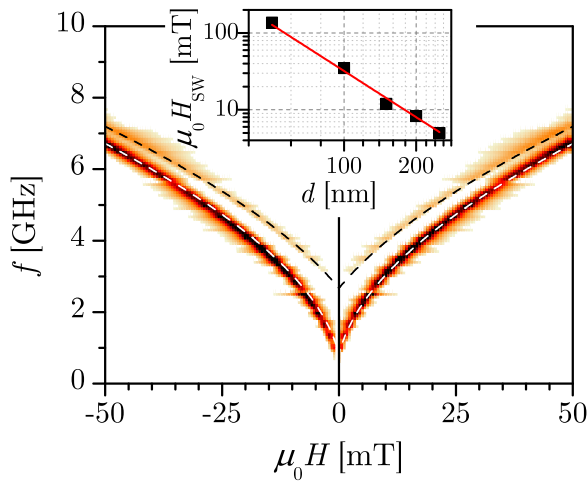}
    \caption{\label{fmr}(Color online) Energy absorption in the broadband FMR experiments as a function of both magnetic field and
    frequency for the 200~nm thick sample. Lines are fits to the theory for the main absorption line (lower curve)
     and for the first spin-wave mode (upper curve). The inset shows the dependence of the spin wave field of the first mode,
     $H_{SW}$, on the sample thickness, $d$, (squares) together with the fit to the expected theoretical $C /d^2$ behavior (solid line).}
    \end{figure}

To determine the exchange constant of Py independently from the magnetization measurements, we performed FMR
experiments on the films from the S2 series. Fig.\ref{fmr} shows the dependence of the energy absorption as function of
magnetic field and frequency for the 200~nm thick sample as a color map.
The spectrum shows the main FMR mode corresponding to the homogeneous excitation of the film that fits well the
expected field dependence given by the Kittel\cite{kittel} equation (white dashed line)
\begin{equation}
f= \frac{\mu_0 \gamma}{2 \pi} \sqrt{H \,(H+ M_s) },
\end{equation}
$\gamma$ being the gyromagnetic ratio. The results for the main mode are very similar in all samples. The value of
$\mu_0 M_s \simeq 1080 \pm 30$~mT extracted from these fittings agrees nicely with the one obtained from SQUID
magnetometry.

Next to the main absorption line, a second resonance (fitted by the black dashed line in Fig.\ref{fmr}) appears in the
spectrum corresponding to the first discrete spin wave (SW) mode associated with the thickness of the sample, $d$. In
this case the frequency dependence follows the expression
\begin{equation}
f= \frac{\mu_0 \gamma}{2 \pi} \sqrt{(H+H_{SW})(H+H_{SW}+ M_s) },
\end{equation}
where $H_{SW}$ is the spin wave field which for the first mode is $H_{SW}= \frac{2A}{M_s} (\pi/d)^2$. The distance in
frequency between the main mode and the first SW mode obviously depends on the sample thickness. We can use this
dependence to obtain the exchange stiffness constant, $A$, in our samples. The inset of Fig.\ref{fmr} shows $H_{SW}$ as
a function of the sample thickness, $d$, together with the fit to the $H_{SW} = C / d^2 $ dependence. From this fitting
we extract a value of $A \simeq (13 \pm 1)\times 10^{-12}$ J/m, which agrees with the usual values for Py.

\begin{figure}[ht]
    \centering
    \includegraphics[width=1.0\linewidth]{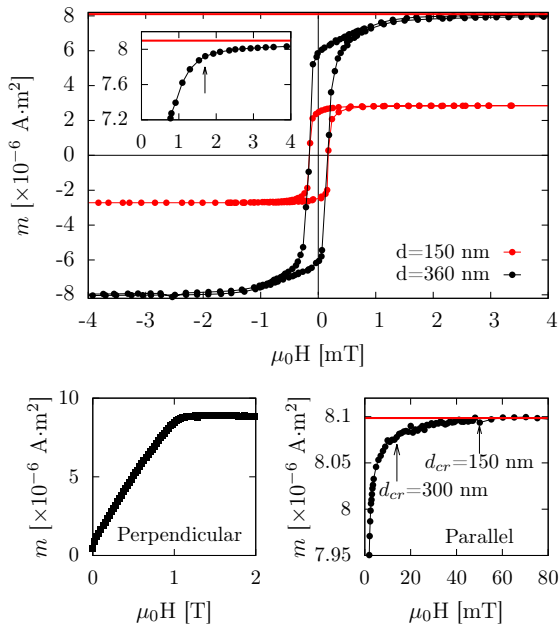}
    \caption{\label{mh}Top panel: magnetic moment versus field applied in plane, for Py films of 150~nm (red
squares) and 360~nm (black circles). The inset shows a zoom of the 360~nm curve. Bottom right: close-up of of the 360~nm curve,
showing at which value of the in-plane field the magnetization saturates ($H^\parallel _{sat}$). Bottom left: measurement
performed on the 360~nm thick sample with the field applied out of plane, in order to determine $H^{sat}_{\bot}$.}
    \end{figure}

The value of $K_{\bot}$ can be estimated using the following relations, which are valid in the case of weak out-of-plane
anisotropy\cite{murayama}:
\begin{align}
&H^{sat}_{\parallel}=2K_{\bot}/\mu_0 M_s\label{hsat}\\
&H^{sat}_{\bot}=M_s[1-2K_{\bot} /(\mu_0 M_s^2)] \label{hsat2}
\end{align}
where $H^{sat}_{\parallel}$, $H^{sat}_{\bot}$ are the fields at which saturation is reached when the field is applied
respectively parallel or perpendicular to the film plane, and $M_s$ is the saturation magnetization. By determining
$H^{sat}_{\parallel}$ and $H^{sat}_{\bot}$ from the magnetization loops, $M_s$ and $K_{\bot}$ can be estimated. In
Fig.\ref{mh} we show magnetic hysteresis loops for unpatterned films of different thicknesses, with the field applied
parallel to the film plane.
The top graph shows the measurements for 150~nm (red squares) and 360~nm (black circles) thick samples from the S2
series. The 360~nm thick sample shows roughly linear decrease of the magnetization between the saturation field and the
remanent field, which is a typical signature of the presence of stripe domains. From this curve, $\mu_0
H_\parallel^{sat}$ is estimated to be about 2 mT. $\mu_0 H^{sat}_{\bot}$, determined from the bottom left panel which
shows a zoom of the $M(H)$ loop of the same sample but with the field applied perpendicular to the film plane, is about
1000 $\pm$ 200 mT. These values lead to an anisotropy $K_\bot \simeq (8.0 \pm 0.4) \times 10^2$ J/m$^3$, which,
combined with the value of $A$, gives $d_{cr} \simeq$ 800~nm, well above the experimental observation. However, by
looking more closely to the magnetization curve when the field is in-plane (a zoom is showed in the bottom right
panel), we can notice that at 2~mT the magnetization is not fully saturated: the value keep on increasing with a much
lower slope and the saturation of 8.1$\cdot$10$^{-6}$ A$\cdot$m$^2$ is reached at around 50~mT. The variation of the
magnetization value between 2~mT and 50 mT is very small and could be due either to the effect of the out-of-plane
anisotropy or to trapped magnetic moments getting aligned or both. 50~mT sets the maximum for the possible values of
$\mu_0 H_\parallel^{sat}$ (minimum $d_{cr}$). With this value, the critical thickness results to be about 150~nm
($K_\bot \simeq (2.1 \pm 0.4) \times 10^4$ J/m$^3$), which is lower than what obtained from MFM measurements. A
$d_{cr}$ of 300~nm, can be obtained if $\mu_0 H_\parallel^{sat} \simeq$ 14~mT ($K_\bot \simeq (5.6 \pm 0.4) \times
10^3$ J/m$^3$), that is compatible with the magnetization data. Interestingly, the curves for the samples 200 and
250~nm thick of the S2 series (not shown here) also show a linear decrease down to remanence, even if less pronounced.
As mentioned above this behavior is a signature of stripe-like magnetic domains, but for this range of thicknesses no
clear stripes are observed with MFM.

The value of $d_{cr}$ was estimated for this particular set of samples of the S2 series. However, a change in the
deposition conditions can influence the magnetic properties of Py (especially $K_{\bot}$), which results in a different
value for $d_{cr}$. In general, negligible differences are expected amongst samples prepared in the same deposition
system. However, changes to the setup which influence the deposition rate or the magnetic configuration inside the
chamber can lead to a variation of $d_{cr}$.Therefore $d_{cr}$ is not to be taken as an exact value, but as an
approximate value of the thickness where to expect
the appearance of stripes. For our discussion we will consider a $d_{cr}$ value of about 300~nm \\
Another point to note is that the numbers confirm that we are dealing with the weak stripe regime. Defining the quality
factor Q = 2$K_{\bot}$/($\mu_0 M_s^2$), we find Q $\approx$~0.05. Note that $\mu_0 M_s^2 / 2$ is sometimes called
$K_d$, the stray field energy coefficient. Strong stripes, where the magnetization direction remains perpendicular to
the surface for all values of the film thickness occur for Q
> 1~\cite{magn_domains,virot12}, and our films are clearly far from that regime.

\subsection{Magnetoresistance measurements}\label{secmr}
Magnetometry and MFM measurements suggest the presence of a non homogeneous magnetization in a large thickness regime
below the appearance of stripes. For the samples in this regime, the magnetic curves show a linear behavior and MFM
images for patterned samples do indicate the presence of an out-of-plane magnetic component, resulting in
cross-tie-like domain walls.

To gain more insight we performed magnetoresistance (MR) measurements on 10 $\mu$m wide bridges. As shown in
Sec.\ref{secmag}, the confinement does not affect the presence of stripes. Moreover, characterization of the relation
between resistance and magnetic configuration in patterned samples can become useful when Py has to be combined with
other layers in devices such as S/F/S junctions. For this reason all measurements were taken at low temperature (5~K).
\begin{figure}[ht]
\begin{center}
\includegraphics[width=0.9\linewidth]{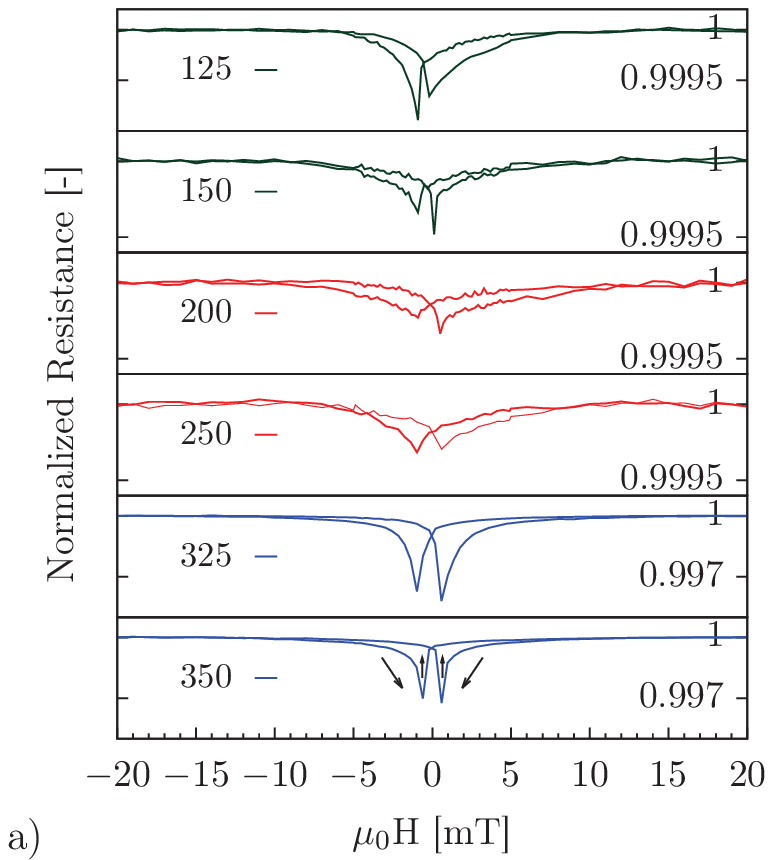}
\includegraphics[width=0.9\linewidth]{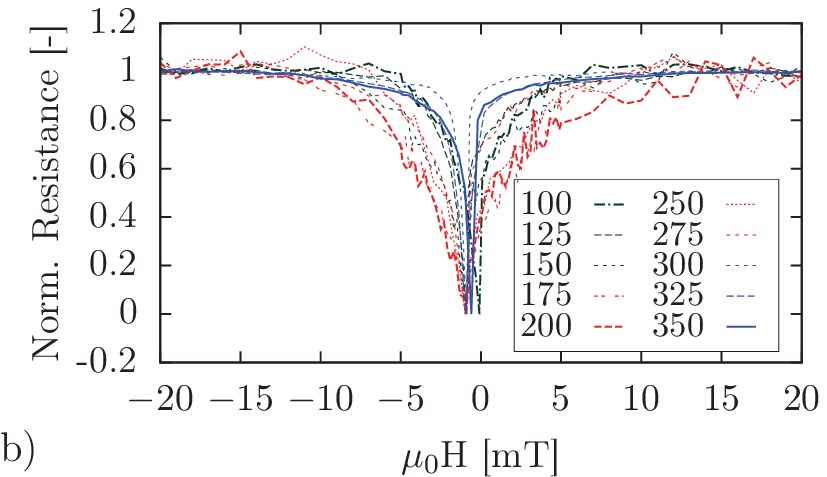}
\caption{\label{mr1} (Color online)(a) Magnetoresistance (MR) traces for patterned Py films (10~$\mu$m wide bridge), for different thicknesses denoted on the left
(in nm). The curves are normalized by $R$(-20~mT). Note the change of scale when going from 250~nm to 325~nm thick structures.
(b) all traces are shown together, normalized by $R$(-20~mT) and the dip height.
All measurements were taken at 5~K. In the bottom plot of (a) the arrows show the sweeping direction of the two curves; see details in the text.}.
\end{center}
\end{figure}
\begin{figure*}[ht]
\includegraphics[width=0.9\linewidth]{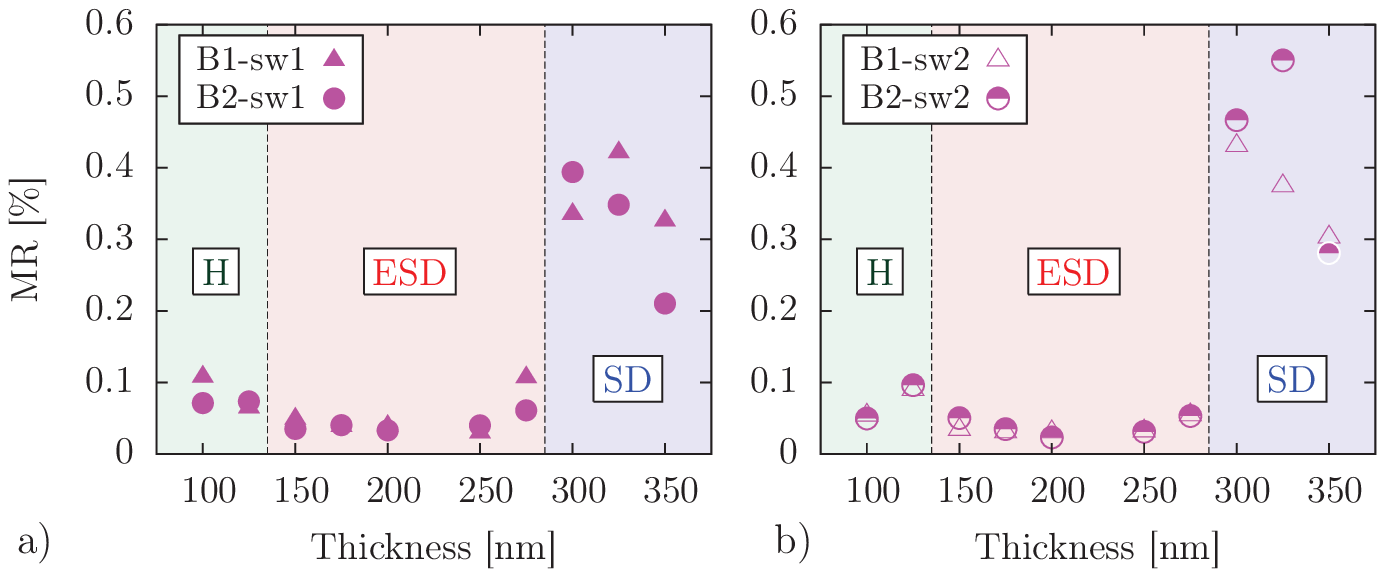}\\
\includegraphics[width=0.9\linewidth]{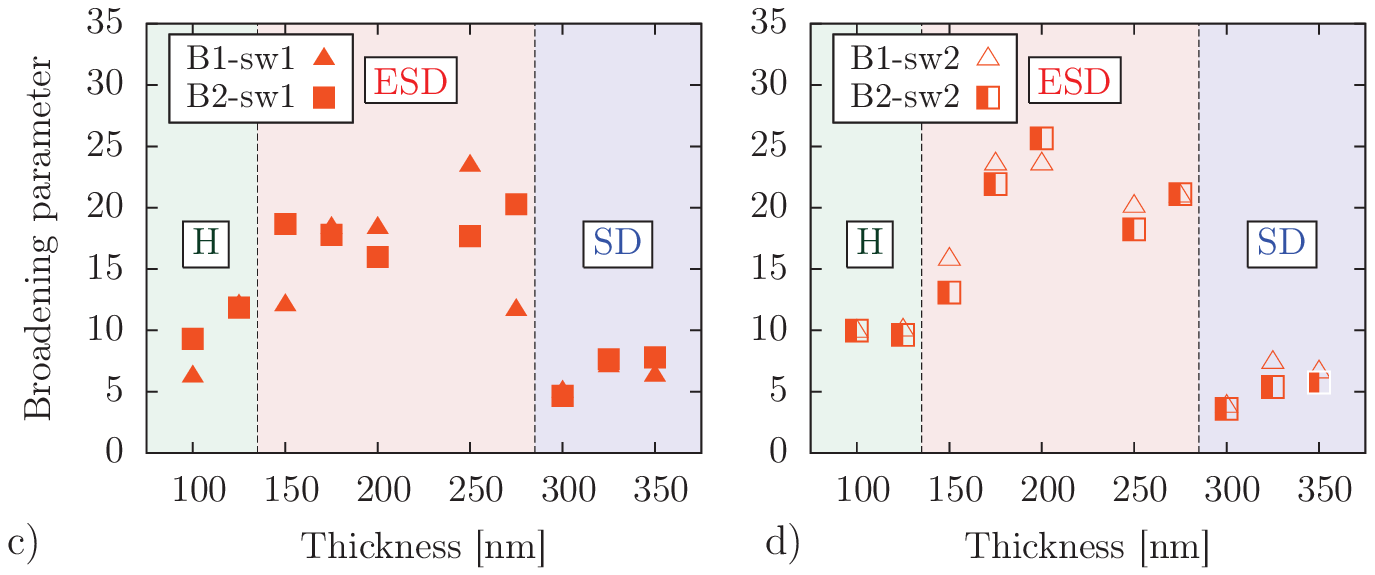}
 \caption{\label{mr2} (Color online) Thickness dependence (a,b) of the magnetoresistance (MR) ratio and (c,d) of the broadening parameter
 for Py films patterned in 10~$\mu$m wide bridges obtained from the $R(H)$ curves given in
Fig.\ref{mr1}. Two bridges (denoted B1 and B2) were patterned on each film. Panels (a) and (c) show the data obtained
from the backward magnetic sweep (from positive to negative fields), for both structures; (b) and (d) are for the
forward sweep (from negative to positive fields), for the same structures as in (a) and (c). The MR ratio is defined as
$100\cdot (R_{-20}-R_{min})/R_{min}$ (with $R_{-20}$ = $R$(-20~mT)) and $R_{min}$ the minimum resistance value); the
broadening parameter is the value of the area enclosed by the MR dip of the $R(H)$ curves, after they are normalized by
$R$(-20~mT) and dip height. All the measurements are at 5~K. The dashed vertical lines divide the data in the three
thickness regimes suggested by the measurements: homogeneous (H), emerging stripe domains (ESD) and stripe domains
(SD).}
\end{figure*}
In Fig.\ref{mr1}(a) we show two $R(H)$ curves for each of the three thickness regimes, normalized by the resistance
value at -20~mT ($R_{-20}$). The field is here applied in-plane and parallel to the current direction (longitudinal
configuration). In order to make sure that the SDs were formed, we applied a high field (typically 1.5~T) along the
bridge before starting each magnetic sweep. The same procedure was followed for all samples, also for the thicknesses
where we did not expect stripes. As expected, the curves show a positive anisotropic magnetoresistance and hysteretic
behavior with a switch of the resistance corresponding to the coercive field. It is important to note the different
scale of the y-axis for the thicker samples (325, 350~nm): for these samples the magnetoresistance ratio is one order
of magnitude higher than for the samples in the other two regimes.
This large increase of MR ratio while passing from the ESD to the SD regime, is highlighted in Fig.\ref{mr2}(a,b),
where the value of the MR ratio is plotted versus Py thickness for all the samples of the series. We defined the MR
ratio as $100\cdot(R_{-20}-R_{min})/R_{-20}$, with $R_{min}$ the resistance value of the minimum of the curve. Left and
right panels show the values obtained from the backward and forward sweep, respectively. Two bridges (denoted B1 and
B2) were patterned on each film and the values for both structures of the same sample are shown together in each panel.
The plots show a sharp transition in MR ratio between 275~nm and 300~nm.

Another interesting feature observed in the curves of Fig.\ref{mr1}(a) is the width of the MR dip, which is larger for
the curve in the intermediate regime. The difference is more visible in Fig.\ref{mr1}(b), where all the curves of the
measured series are plotted, normalized by $R_{-20}$ and the dip height. In this way all the dips have the same height
and their shape can be directly compared. From this graph is evident that the curves of the intermediate thickness
regime are broader compared to thicker and thinner samples. To quantify this change in shape of the MR curves we define
a \emph{broadening parameter}, $B_{br}$, given by the area enclosed by the normalized curves of Fig.\ref{mr1}(b). The
results are summarized in Fig.\ref{mr2}(c,d), where the values of $B_{br}$ are presented for the same structures and
sweeps of Fig.\ref{mr2}(a,b). The graphs show that there is a clear broadening of the MR curve (higher value of
$B_{br}$) in the intermediate regime. The broadening sets in at a thickness of about 150~nm, which interestingly enough
is the value of the estimated $d_{cr}$, and decreases in between 275~nm and 300~nm, in conjunction with the strong
increase of the MR ratio.

The combination of Fig.\ref{mr2}(a,b) and (c,d) makes us identify three different magnetic regimes (in the plots
separated by dashed vertical lines and different background colors): the first for thicknesses below approximately
150~nm, the second between 150~nm and 280~nm and the third one above 280~nm, respectively called homogeneous (H),
emerging stripe-domains (ESD) and stripe-domains (SD) regime. For the H regime, as expected, the weak out-of-plane
anisotropy does not play a role and the magnetization is homogeneously in-plane. At around 150~nm we have a change in
the magnetic configuration signaled by a broadening of the MR curve and the appearance of a linear behavior in the
$M(H)$ loops, even if well defined stripes are not developed yet. A second abrupt transition is observed between ESD
and SD regime: the MR ratio increases by one order of magnitude, at the same time the broadening returns to a low
value. Above this threshold the standard SDs, as known from the literature, are also observed in the magnetic
measurements. The existence of an intermediate non-homogeneous state could also explain the data of
Ref.\onlinecite{ramos}, where the FMR spectra for the intermediate thickness (sample S1) shows a peculiar double peak
feature, while MFM and $M(H)$ do not show any signature of inhomogeneity. At this moment we can only speculate on the
precise nature of the ESD regime. It is clearly characterized by the absence of long range order in the perpendicular
component of the magnetization. Looking back at Fig.\ref{sketch}, this could be either considered as an extended domain
wall, or as a state in which the perpendicular components of the magnetization are not ordered yet. Local probes of the
magnetization, such as with polarized neutrons, may shed more light on the nature of the ESD.

\section{Micromagnetic simulations}\label{secsim}
\begin{figure}[ht]
\includegraphics[width=0.9\linewidth]{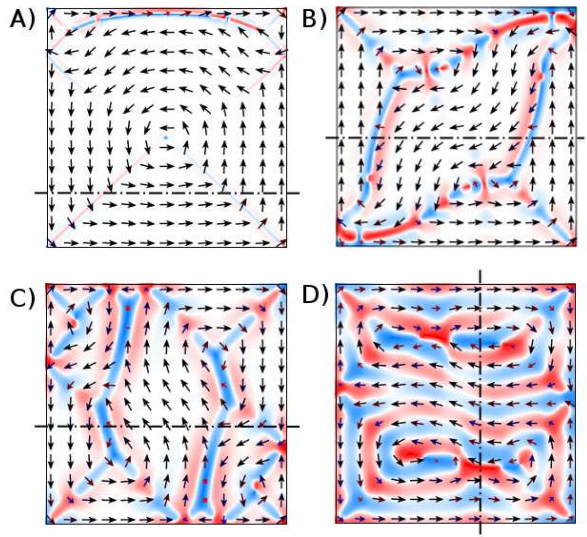}
\caption{\label{sim} (Color online) Micromagnetic simulations (\textsc{oommf} software package) for square structures $4 \times 4~\mu$m$^2$
with different thickness: 100~nm (A), 225~nm (B), 285~nm (C) and 345~nm (D). The images show the magnetization of the $xy$-plane
(film plane) at half of the thickness. The arrows represent the direction of the magnetization in the $xy$-plane, while the color
scale is for the magnetization component along $z$ (red~$+z$, blue~$-z$). In (A), out-of-plane components present along the diagonals
of the square are barely visible. Dot-dash lines indicate the position of cross-sections plotted in Fig.\ref{sim-cross}.}
\end{figure}

\begin{figure}[ht]
\includegraphics[width=1.0\linewidth]{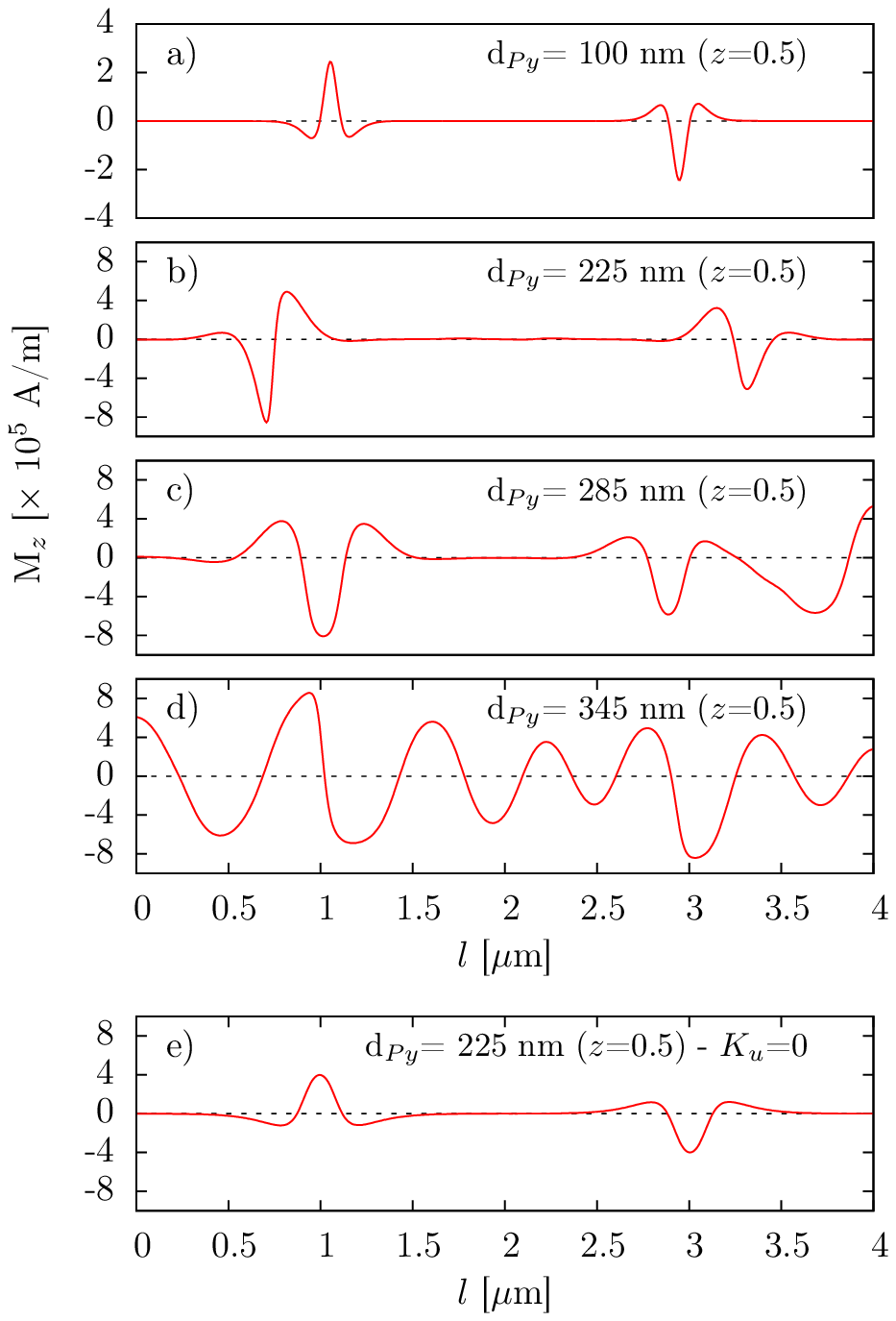}
\caption{\label{sim-cross} (Color on-line)(a,b,c,d) Cross-sections of the images in Fig.\ref{sim} showing the perpendicular
component of the magnetization, along the length of the section $l$. Position and direction of the cross-section are shown
in Fig.\ref{sim}. The origin of the coordinates of the squares is the bottom-left corner. (e) Cross-section for the same
structure as in (b) but with no perpendicular anisotropy.}
\end{figure}

The results of Sec.\ref{secmfm} and \ref{secmr} suggest that in the ESD regime the magnetization is not as homogeneous
as one would expect. In order to better characterize this intermediate regime, we performed micromagnetic calculations
by using the \textsc{oommf} software package~\cite{oommf}. We simulated confined structures, in particular squares $4 \times
4~\mu$m$^2$, with thicknesses in the three regimes: 100, 225, 285 and 345~nm. The parameters used for exchange
stiffness and saturation magnetization are the ones obtained from FMR experiments (see Sec.\ref{secmag}), namely $A =
13 \times 10^{-12}$~J/m and $M_s = 8.59\times 10^{5}~$A/m ($\simeq 1080$~mT). For the uniaxial (out-of plane)
anisotropy we chose $K_\bot = 5.6 \times 10^3$~J/m$^3$, the value extrapolated from MFM measurements
(Sec.\ref{secmag}). In Fig.\ref{sim} we present the results, which show the magnetization in the middle plane of the
sample, that is the $xy$-plane at half of the thickness.
The out-of-plane component of the magnetization is represented by the color scale from red ($M_z$>0) to blue ($M_z$<0),
while the arrows show the direction of the in-plane magnetization. For ease of comparison, Fig.\ref{sim-cross}a, b, c
and d show the behavior of the out-of-plane magnetization when taking a cross-section along a line of the square
structures, as shown in Fig.\ref{sim} (dot-dash lines). The simulations quite accurately reproduce the magnetic
configurations observed with MFM on $10 \times 10~\mu$m$^2$ squares as presented in Fig.\ref{mfm2}. For the thinnest
structure in Fig.\ref{sim}A (100~nm, to be compared with Fig.\ref{mfm2}A) four closure domains are visible, divided by
the diagonals of the square, where an out-of-plane component of the magnetization is barely visible although it shows
up in the cross-section in Fig.~\ref{sim-cross}a. Stripe domains appear in the structure 350~nm thick (Fig.\ref{sim}D,
cf. Fig.\ref{mfm2}E). The stripe width is of the order of 300~nm, in agreement with the experimental value.
Interestingly, the simulation qualitatively reproduce the domain structure found in the ESD regime: in Fig.\ref{sim}B
the closure domains are smaller than in Fig.\ref{sim}A, and two types of walls appear, both of which are also visible
in Fig.\ref{mfm2}C : ``broken'' domain walls which show a sequence of red-blue equal to up-down magnetization
directions; and walls which consist of an up and a down component running parallel to each other and separate closure
domains with anti-parallel in-plane magnetization. Such walls, with an up and a down component, are known as asymmetric
Bloch walls. The rotation of the magnetization within these domain walls, also observed in MFM, is very similar to what
happens in a wall between two stripe domains, with the difference that in the stripes the in-plane magnetization is
parallel. As a comparison, in Fig.\ref{sim-cross}e we show the cross section for a 225~nm thick structure,  with no
perpendicular anisotropy. In this case the simulation was run with a randomized in-plane anisotropy (with $K$ =
100~J/m$^3$) and the cross-section is at same position as in Fig.\ref{sim}A. By comparing it with Fig.\ref{sim-cross}b
we can notice that, without perpendicular anisotropy, the domain wall configuration is the same as in the homogeneous
regime (cf. Fig.\ref{sim-cross}a) and the amplitude of the out-of-plane $M_z$ component is significantly lower than in
Fig.\ref{sim-cross}b. From Fig.\ref{sim}C we can see that, by increasing the thickness further to 285~nm, the domain
walls are stretched but, because we are still below $d_{cr}$, stripe domains are not formed yet. 

\section{Discussion and conclusions}\label{seccon}
Summarizing, we studied patterned and unpatterned films of Py with different thicknesses, below and above the critical
value $d_{cr}$ for the appearance of stripe domains. Magnetoresistance measurements, combined with MFM and SQUID
magnetometry suggest the existence of three magnetic regimes: homogeneous (H), emerging stripe domains (ESD) and stripe
domains (SD). More quantitatively, their appearance can be characterized by the dimensionless parameter $\tilde{d}$ =
$d / \sqrt{A/K_d}$, which allows a comparison with earlier work. With the numbers given before, we have $\tilde{d}$(100
nm) = 10, $\tilde{d}$(225 nm) = 22.5, and with $d_{cr}$ = ($2 \pi / \sqrt{Q} ) \sqrt{A / K_d}$ it follows that
$\tilde{d}_{cr}$~= $2 \pi / \sqrt{Q}$~= 28. In the H regime, up to $\tilde{d}$~=10, there is no evidence of stripes in
films or strips, the magnetization is fully in-plane,  and in confined structures domain walls are of the simple Bloch
type. In the SD regime, above $d_{cr}$, the stripes are well developed (as it clearly appears from MFM measurements)
and they are signaled by the peculiar shape of the $M(H)$ loops as well as from an increase of the MR ratio. In the ESD
regime between $\tilde{d}$~=10 and $\tilde{d}$~= 28 (from about 0.5~$d_{cr}$ to $d_{cr}$), stripes are not visible in
MFM images but both strips and square structures easily becomes less homogeneous. This is signaled by the peculiar
domain walls observed with MFM, which have a stronger out-of-plane component, by a linear behavior in $M(H)$ loops and
by a broader dip characterizing the MR curves. In this regime the MR ratio is still much smaller than in the SD regime.
Micromagnetic simulations for the squares reproduce the configuration of magnetic domains and domain walls in all three
regimes quite well. In particular they show how in the ESD regime the perpendicular anisotropy leads to a richer domain
wall configuration, especially in confined structures where the influence of demagnetizing field is weaker than in
films. This might not have been expected from the phase diagram for domain wall types given in Refs.\cite{magn_domains}
and\cite{ramstock96}, obtained from a two-dimensional calculation for a strip of fixed width/thickness ratio 4~:~1. In
that case, asymmetric Bloch walls were found for $\tilde{d}$~> 7, in other words no changes occurred for the behavior
up to $d_{cr}$. For our strips, the width/thickness ratio is significantly larger, which may explain the
difference. \\
Concluding, we have shown how in particular the magnetoresistance evolves of Py films below the onset of the magnetic
stripe phase. A strong change in MR is found at the critical thickness, while well below $d_{cr}$ the MR dips show
significant broadening. Micromagnetic simulations show good agreement with MFM measurements on confined structures, and
emphasize the difference between such structures and long strips.
\section{Acknowledgments}
Technical support from D. Boltje and M.B.S. Hesselberth and help from Annette Mense in the early stage of the project
are gratefully acknowledged. This work is part of the research programme of the Foundation for Fundamental Research on
Matter (FOM), which is part of the Netherlands Organisation for Scientific Research (NWO). The work was also supported
by the EU COST action MP1201 'NanoSC'. J.M.H. and A.G.-S acknowledge support from Universitat de Barcelona and from the
Spanish Government Project No. MAT2011-23698.


\end{document}